\documentclass[%
 reprint,
 amsmath,amssymb,
 aps,
prb,
]{revtex4-2}

\usepackage{graphicx}% Include figure files
\usepackage{dcolumn}% Align table columns on decimal point
\usepackage{bm}
\usepackage{amssymb}
\usepackage{gensymb}

\begin{document}

\preprint{APS/123-QED}

\title{A pathway to improve Nb$_3$Sn for SRF accelerator cavities via Zr inclusion }

\author{M.Tripathi$^{1}$}
\author{M. S. Shakel$^2$ }
\author{J.-Yel Lee$^1$}
\author{H. E. Elsayed-Ali$^2$}
%\author{X. Su$^1$}
\author{S. Posen$^1$}
\author{G. V. Eremeev$^1$}
\affiliation{$^1$Fermi National Accelerator Laboratory, Batavia, IL-60510, USA}
\affiliation{$^2$Department of Electrical and Computer Engineering, Old Dominion University, Norfolk, VA 23529, USA}

\date{\today}

\begin{abstract}
Nb$_3$Sn has the potential to exceed the performance of niobium-based superconducting radio-frequency (SRF) cavities, both in particle accelerators and in high-$Q_0$ cavities operating in multi-tesla magnetic fields for axion dark matter searches. We studied how the incorporation of Zr into the Nb$_3$Sn matrix by co-sputtering affects the microstructure and superconducting properties of Nb$_3$Sn films. As the atomic fraction of Zr increased, the surface morphology improved and the bulk and surface void densities decreased. At the optimal Zr content, the superconducting properties improved and a thinner surface niobium oxide formed. These results identify co-sputtered Nb$_3$Sn:Zr as a promising material to advance the performance limits of SRF cavities.
\end{abstract}

\maketitle

\section{\label{sec:level1}Introduction}

%Superconducting technology and materials development have transformed the operation and design of particle accelerator design and operation. 
Contemporary particle accelerators: the LHC (CERN), LCLS-II HE (SLAC), and the future PIP-II (Fermilab), use Niobium (Nb) based superconducting radio-frequency (RF) cavities to accelerate particle beams at high acceleration gradients while generating very small power dissipation. After several decades of advances in preparation methods, state-of-the-art Nb-based superconducting RF cavities can achieve accelerating gradients of $\sim 50~\mathrm{MV/m}$ with quality factors ($Q_{\mathrm{0}}$) $>10^{10}$ while operating at $2~\mathrm{K}$\cite{Padamsee2017}. At this stage, the performance levels and the scope of further improvement are mainly limited by the fundamental superconducting properties of Nb~\cite{Padamsee2008,belomestnykh2022key,padamsee2020history}. For future accelerator projects, such as FCC-ee, would require enhanced accelerator performance and significantly reduce operational costs~\cite{venturini2023progress,aull2018material,Delsolaro2023}. Another application of superconducting radio frequency is in the field of dark matter searches, where RF cavities can provide very high sensitivity and scan rates for detecting the extremely small signals produced by axions~\cite{Graham2015,Braine2020}. As the scan rate of a haloscope grows with the cavity quality factor, SRF cavities that retain a high $Q_0$ under multi-tesla fields can substantially accelerate these searches~\cite{Posen2023}fields.

Current research \& development is focused on exploring alternative superconducting materials that can push the current limits of accelerating gradients and $Q_0$, reduce the cryogenic load by operating at higher temperatures, and maintain high $Q_0$ values even in the presence of strong magnetic fields. Nb$_3$Sn superconductor has emerged as the most promising candidate, attributing to its twice as large superconducting transition temperature ($T_C \sim 18~\mathrm{K}$) and super-heating field ($\mu_0 H_{sh} \sim 425~\mathrm{mT}$) in comparison of elemental Nb ($T_c \sim 9~\mathrm{K}$; $\mu_0 H_{sh} \sim 220~\mathrm{mT}$). The factor of two improvements in $T_c$ has the potential for a two-order reduction in RF losses and a factor of two increase in the maximum acceleration gradient~\cite{Padamsee2008,AMValentino2016}. On the other hand, high $H_{sh}$ of Nb$_3$Sn would allow high energy linear colliders to reach their design energy with far fewer cavities~\cite{Kneisel2003,Posen2017}. Theoretically, Nb$_3$Sn/Nb structures are predicted to reach accelerating gradients as high as $\sim 100~\mathrm{MV/m}$ while operating at $4~\mathrm{K}$~\cite{Posen2022Snowmass}. Experimentally, however, the best demonstrated value to date is $\sim 24~\mathrm{MV/m}$~\cite{Posen2021,Lechner_2025}. This difference between the predicted and experimentally achieved values is mainly attributed to the quality of the material: defects such as Sn-deficient regions, thin grains and Sn segregation at grain boundaries, together with surface roughness, act as preferential sites for vortex penetration and lead to premature quench~\cite{Carlson2021,Lechner_2025,Kubo2013}.

A promising strategy for material optimization is the incorporation of a small amount of a third element into the Nb$_3$Sn A15 structural phase either as a ternary dopant or as an interstitial defect. Certain elements may act as beneficial defects that increase grain-boundary density and enhance Sn diffusion during synthesis, thereby promoting the formation of a Sn-rich, homogeneous, and uniformly stoichiometric Nb$_3$Sn A15 phase while preserving the T$_C$ and the density of states at the Fermi level below T$_C$ \cite{kelley2021ab, tarantini2016significant, tarantini2021origin}. This approach is widely used in Nb$_3$Sn-based superconducting magnet development, where elemental inclusion at interstitial sites enhances magnetic flux pinning efficiency and the critical current density \cite{XU2021, LEE2024176044}. However, this approach remains relatively less explored for Nb$_3$Sn-based RF accelerator cavities, primarily because methods have not been developed for introducing a controlled ternary incorporation during the deposition of Nb$_3$Sn on complex shaped cavities. Recent advances in cylindrical magnetron deposition techniques for RF cavities \cite{shakel2023commissioning} enable the use of multiple targets with complex elemental configurations and tunable deposition parameters, allowing fine control over the stoichiometry of deposited materials on complex shaped cavity surfaces. This method provides a practical route for introducing additional elements during Nb$_3$Sn growth on RF cavities.

In this work, we studied how Zr incorporation into Nb$_3$Sn via co-sputter deposition influences its structural and superconducting properties, with the aim of identifying correlations relevant to the RF performance of Nb$_3$Sn-coated cavities for both particle accelerators and high-field axion searches. We fabricated Nb$_3$Sn thin films with different atomic fractions of Zr by simultaneous co-sputtering of elemental Nb, Sn and Zr using a magnetron sputtering setup, and characterized their cross-sectional and surface microstructure, superconducting properties, and surface and bulk chemical composition.

%Here, our aim is to understand how Zr inclusion in Nb$_3$Sn via a co-sputter deposition method can influence the structural and superconducting properties which can correlate to the RF performance of cavities.  effect the route for effectively reducing the intrinsic defects in magnetron sputter coated Nb$_3$Sn SRF cavities and eventually improve the overall RF performance. With this motive, we have fabricated Nb$_3$Sn thin film samples with different atomic fractions of Zr by simultaneousness co-sputtering of elemental Nb, Sn and Zr using a magnetron sputtering setup. We used various material characterization techniques to explore how Zr inclusion in Nb$_3$Sn sputtered material impacts the micro-structural, superconducting and flux pinning properties of Nb$_3$Sn.

The results of this study provide a material level insight into the effects of Zr on the state-of-art sputter coated Nb$_3$Sn films and offer a new pathway to tailor RF performance of SRF cavities and  will also be helpful for the ongoing hunt of ideal platform for axion dark matter research.  
%The results of this study will provide a material level advancement in the state-of-art sputter coated Nb$_3$Sn and will offer a new pathway to improve RF performance of SRF cavities. 
%Additionally, it is worth emphasizing that this effort can also benefit the development of axion haloscopes for dark matter searches. Axion haloscopes can provide very high sensitivity and scan rates for detecting the extremely small signal produced by the axions\cite{Graham2015,Braine2020}. The search sensitivity and scan rate of axion haloscopes can together be improved by enhancing the effective quality factor (Q$_{eff}$) of superconducting RF cavities immersed in multi-Tesla magnetic field\cite{Posen2023}. Nb$_3$SnZr-coated cavities have an advantage of Zr in Nb$_3$Sn as micro-structural heterogeneities, which can increase the super-current flux-pinning efficiency and hence enhance Q$_{eff}$ in high magnetic fields. Therefore, this study will also be helpful for the ongoing hunt of ideal platform for axion dark matter research. 

\section{Sample preparation and  Experimental methods}
Nb$_3$Sn films with different Zr concentrations were grown using an Orion-5 magnetron sputtering system (AJA International, Inc., USA). Nb and Sn sputtering targets, with dimensions of 2.00$^{\prime\prime}$ diameter $\times$ 0.25$^{\prime\prime}$ thickness and 99.999\% purity, were obtained from Kurt J. Lesker. The Zr target, with the same dimensions and purity, was obtained from ACI Alloys, Inc.
The Nb target was sputtered by a DC power supply while the Sn and Zr targets were sputtered by RF power supplies. All films were deposited in a high-vacuum chamber with a base pressure below 2.5 × 10$^{-7}$ Torr. Depositions were conducted at 3 mTorr using high-purity Ar gas (99.99\%) with a flow rate of 20 sccm. Prior to depositing the Zr-doped Nb$_3$Sn film, an approximately 200 nm thick Nb barrier layer was deposited on the sapphire substrates (430 $\mu$ m thick, double-sided polished, C-M plane, obtained from University Wafer, Inc.). The Nb$_3$Sn:Zr layer was grown by the simultaneous co-sputtering of Nb, Sn, and Zr.  The Nb and Sn contents were kept constants for all depositions, while the RF power applied for Zr sputtering was varied across depositions to achieve different Zr concentrations in the Nb$_3$Sn films. The total thickness of the co-sputtered Nb$_3$Sn:Zr layer was estimated to be in the range of 1–1.2$\mu m$. For all depositions, the substrate temperature was kept at room temperature, and the substrate holder was rotated at 30 rpm throughout the sputtering process to ensure uniform film deposition.  The substrate holder was rotated at 30 rpm throughout the sputtering process to maintain a homogeneous deposition. After deposition, the samples were annealed at 920$\degree$C for 3 hours in a furnace at a base pressure less than 2x 10$^{-7}$ mbar.
\begin{figure*}
    \includegraphics[width=1\linewidth]{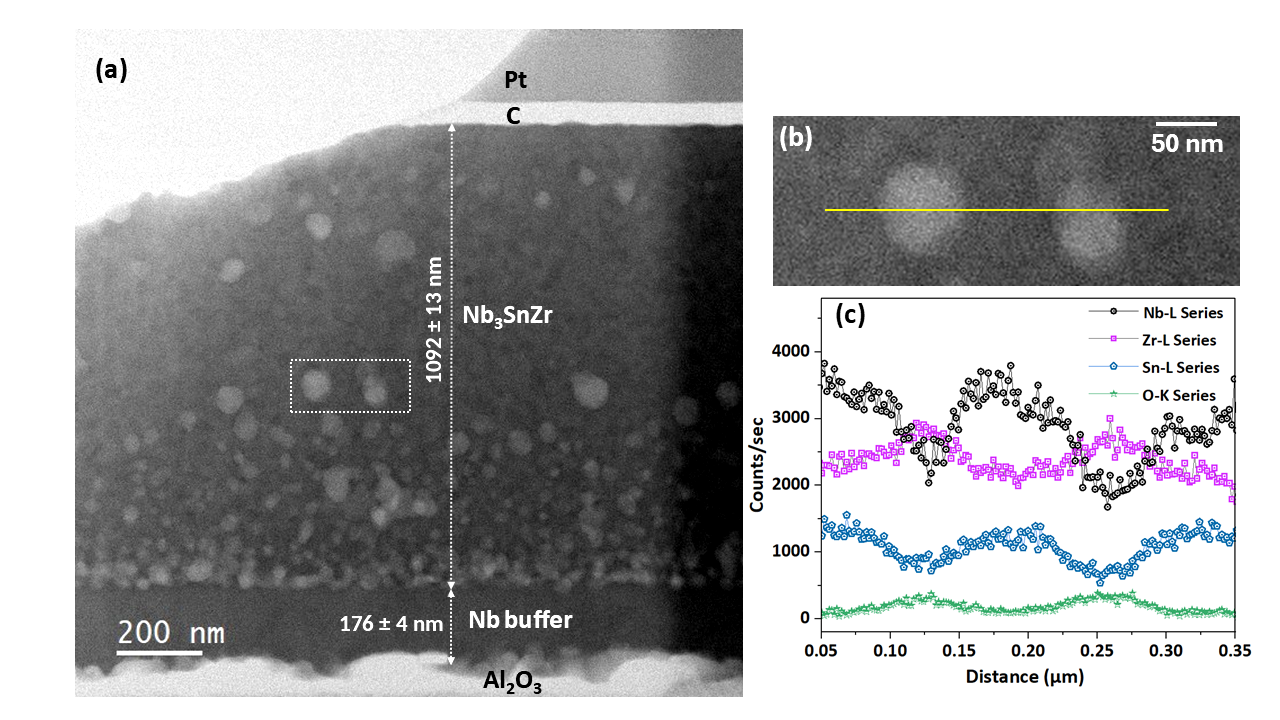}
    \caption{Cross-sectional STEM characterization of a representative Nb$_3$Sn:Zr film with 23.6~at.\% Zr. (a) Bright-field STEM image. (b) Enlarged view of the region marked by the dashed rectangle in (a). The yellow line indicates the path of the EDS line scan. (c) EDS line profiles of the Nb-$L$, Zr-$L$, Sn-$L$ and O-$K$ signals along the line in (b), across the two precipitates.}
    \label{fig:enter-label}
\end{figure*}
The cross-sectional microstructure of samples were studied by Transmission Electron Microscopy (TEM) and High-Angle Annular Dark-Field Scanning Transmission Electron Microscopy (HAADF-STEM). TEM images were acquired using using JEOL JEM-2100 FasTEM microscope using double tilt sample holder under ambient conditions. JEOL ARM200CF Aberration-Corrected TEM microscope was used for HAADF-STEM measurements. The surface morphology of samples was studied by Secondary Electron Microscopy (SEM). SEM imaging and sample preparations for TEM foils were performed using a FEI Helios Nanolab 600 focused ion beam (FIB). The thickness of TEM foils were reduced by Ga ions at 30 kV and 87 pA to 0.4 nA. Afterwards, the samples were polished using 5 kV Ga$^+$ ions at 28 pA, to remove the damaged surface layers of the TEM lamella. The FIB of the same instrument was also utilized to create patterns to define the current channels for electron-transport resistivity measurements. A 9 T PPMS (Quantum Design, Inc.) was used to measure the temperature- and magnetic-field-dependent resistivity and bulk magnetometry. Before all temperature- and field-dependent measurements, the sample temperature was raised to 25 K, well above the superconducting transition temperature of Nb$_3$Sn, and a $\it{de-gaussing}$ protocol\cite{mastrogiacomo2008soft} was applied by oscillating the magnetic field from +9 T to $-$9 T with variable amplitudes. This procedure is reported to reduce the trapped magnetic field to below 10 Oe\cite{design2009using}. All temperature-dependent resistivity measurements were performed in zero applied magnetic field. A customized SPECS-X-ray photoemission spectroscopy (XPS) instrument is used to record core-level Nb, Sn and Zr XPS spectra of different Nb$_3$Sn samples. We used Al K-alpha (E = 1486.3eV) to measure the core level spectrum. Commercially available CASAXPS software package is used for peak fitting of the XPS profiles.  The in-elastically scattered electron backgrounds are simulated using the Shirley baseline model. For X-ray diffraction measurement, we used a Rikagu SmartLab instrument with a 3 kW Cu anode, Ge(220) two bounce monochromator, and a 5-circle diffractometer. The incident Cu K-alpha X-ray flux of wavelength $\lambda$ = 1.541\textup{~\AA} on the sample was 3.3 × 10$^7$ photons per second. The $\chi$ vs $\psi$ measurements are measured across an asymmetric Bragg's reflection of Nb (200).  IONTOF time-of-flight secondary ion mass spectrometry (ToF-SIMS) dual beam system was used for the depth profiling. The liquid bismuth ion beam (Bi+) was used for secondary ion measurements, while the cesium (Cs+) gun of 1 keV energy was used for the sputter erosion during depth profiling.

\section{Experimental results}
\subsection{Incorporation of Zr in Nb$_3$Sn host matrix: Modifications in cross-sectional microstructure and crystal structure}

The incorporation of Zr into the Nb$_3$Sn matrix during simultaneous co-sputtering is addressed first, prior to evaluating its effect on the properties of Nb$_3$Sn. To this end, the cross-sectional microstructure of the deposited films was examined by scanning transmission electron microscopy (STEM). Fig.~1(a) shows a bright-field (BF) cross-sectional STEM image of a representative film containing $\approx$23.6~at.\% Zr.
The film stack consists of a $184 \pm 21$~nm Nb buffer layer grown on an $\alpha$-Al$_2$O$_3$ (001) substrate, followed by a $1084 \pm 13$~nm ($\approx$1~$\mu$m) Zr-doped Nb$_3$Sn layer. The image also reveals several bright precipitates with an average dimension of $57 \pm 14$~nm (range 37--82~nm) and an average projected area of $(2.4 \pm 1.2) \times 10^{3}$~nm$^2$ (range $0.9$--$4.8 \times 10^{3}$~nm$^2$). 
To identify the elemental composition of these regions, energy-dispersive X-ray spectroscopy (EDS) was performed in STEM mode. A region of interest containing two bright regions, marked by the dashed rectangle in Fig.~1(a), is shown enlarged in Fig.~1(b), and the corresponding EDS line scan along the yellow line is displayed in Fig.~1(c). Across both bright regions (at $\sim$0.12 and $\sim$0.25~$\mu$m along the scan), the Nb and Sn signals decrease, while the Zr and O signals increase. The bright contrast of these regions in the bright-field image is also consistent with a phase of lower average atomic number than the surrounding Nb$_3$Sn matrix. Together, these observations indicate that the bright regions are Zr- and O-rich precipitates, most likely zirconium oxide (ZrO$_2$). The Nb, Sn and Zr signals do not vanish within the precipitates, since the electron probe also samples the surrounding matrix through the finite thickness ($\approx$50~nm) of the lamella.

Since Zr has a much higher affinity for oxygen than Nb, i.e., ZrO$_2$ has a more negative Gibbs free energy of formation than the Nb oxides \cite{Jacob2010NbOxides, Balachandran_2019, XU2021}, Zr can form oxide nano-precipitates in the Nb$_3$Sn host matrix through internal oxidation during annealing. The formation of ZrO$_2$ precipitates by internal oxidation has previously been reported in Nb$_3$Sn wires \cite{LEE2024176044, XU2021}, where an oxygen-supplying precursor was deliberately incorporated and served as the primary oxygen source for Zr. In our films, no oxygen source was intentionally added, and we identify two possible sources: (i) native surface oxides, which dissociate at the annealing temperature and release oxygen that diffuses inward through the Nb$_3$Sn layer, and (ii) the Al$_2$O$_3$ substrate, from which oxygen can enter the Nb buffer at the Nb/Al$_2$O$_3$ interface and subsequently diffuse into the Nb$_3$Sn:Zr layer. Oxygen transfer from sapphire into Nb films has been reported at elevated temperatures, where it leads to internal oxidation at the Nb/Al$_2$O$_3$ interface \cite{Eusterholz2023}, and has also been suggested as the origin of oxygen segregated at the surface of Nb films grown on sapphire \cite{Surgers1992}. Residual oxygen incorporated during sputter deposition may provide an additional contribution.

To evaluate the effect of Zr content on the precipitate morphology, cross-sections of films with Zr atomic fractions ranging from 10.9 to 24.8\% and a similar Nb$_3$Sn:Zr thickness (0.9--1.1~$\mu$m) were examined (Fig.~S1 in the Supplementary Material). Increasing the Zr content from 10.6 to 24.8\% increases both the average precipitate dimension, from $40 \pm 18$ to $50 \pm 22$~nm, and the average precipitate area, from $(1.2 \pm 1.0) \times 10^{3}$ to $(2.0 \pm 1.7) \times 10^{3}$~nm$^2$. $^{-6}$). Notably, the number density changes only slightly (133 and 124 µm$^{-2}$), whereas the projected area fraction increases from 16 to 24\%, indicating that the additional Zr is incorporated mainly by growth of the precipitates rather than by additional nucleation. The spatial distribution of the precipitates also depends strongly on the Zr content. At lower Zr content, the precipitates are confined mainly to bands near the top surface and the Nb buffer, leaving the central part of the film almost free of precipitates, whereas at higher Zr content they are distributed throughout the film. Since classical internal-oxidation theory predicts a slower advance of the oxidation front at higher solute content \cite{Wagner1959}, the uniform distribution at higher Zr content suggests that oxygen is also present within the film, in addition to being supplied from the interfaces, for example through incorporation during deposition.

\begin{figure*}
    \includegraphics[width=1\linewidth]{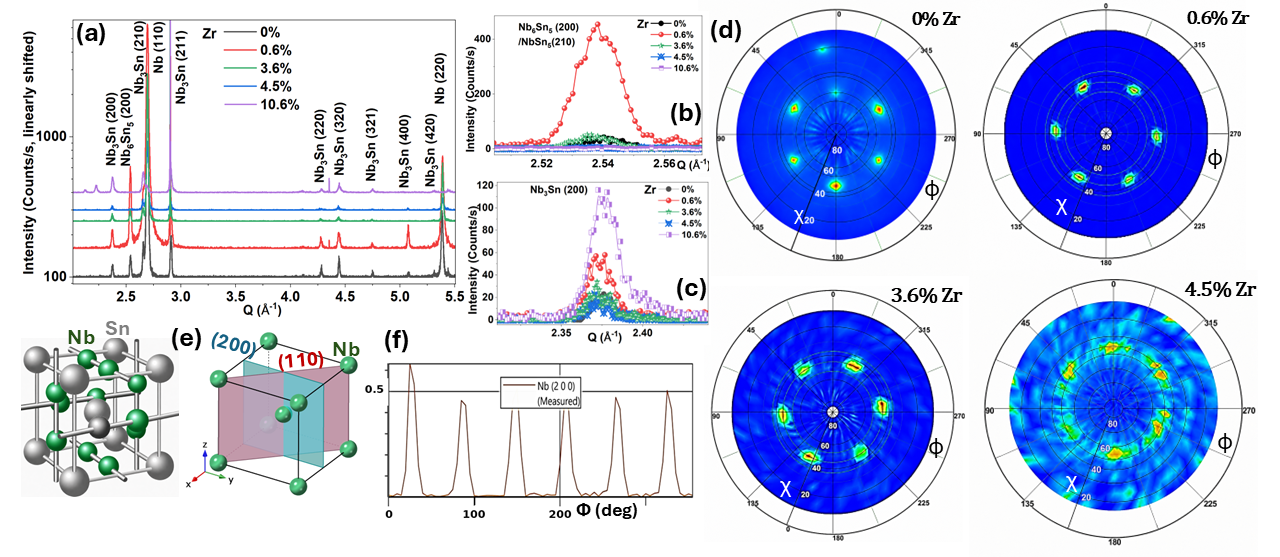}
    \caption{Comparative X-ray diffraction profiles for Nb$_3$Sn thin films with Zr concentration varying from 0 at\%-10.6 at\% : (a) ${2\theta - \omega}$ scans for Nb$_3$Sn thin films with Zr concentration varying from 0 at\%-10.6 at\%, (b) Peak profile of feature attributed to Nb$_{5}$Sn$_{6}$/NbSn$_{2}$ (c) Peak profile of Nb$_3$Sn (200) with variable Zr concentration (d) Pole-figure measurements of Nb$_3$Sn with four different concentration values with respect to Nb $\chi$ (200) plane. (e) Illustration of Nb$_3$Sn and Nb crystal structure along with planer orientations. (f) Intensity distribution of in $\chi-\phi$ scan along Nb (200) direction.}
    \label{fig:enter-label}
\end{figure*}

Fig.~2 elaborates the modification of the crystallographic structure of Nb$_3$Sn with the inclusion of Zr. ${2\theta - \omega}$ X-ray diffraction (XRD) measurements of Zr-doped Nb$_3$Sn samples with Zr concentration ranging from $\approx$ 0-10.6 at\% are presented in Fig.~2(a). Based on the ${2\theta - \omega}$ XRD profiles,  Nb$_3$Sn crystallizes in the A15 structure with the Pm-3n space group and a lattice parameter of $a = b = c = 5.273~\AA$. 

There are three main observations from these measurements: (i) The appearance of two low-angle peaks centered at ${Q = 2.13~\AA^{-1}}$ (2$\theta$ = 30.14\degree) and ${Q = 2.24~\AA^{-1}}$ (2$\theta$ = 31.74\degree) when Zr is increased to ${\approx 10.6 at\%}$. These two peaks can be attributed to the orthorhombic phase of ZrO$_2$ with the Pnma(62) space group\cite{ZrO2}. This observation further supports that, instead of being incorporated into the A15 unit-cell structure of Nb$_3$Sn, the ZrO$_2$ particles form their own phase-separated ordering. Notably, the orthorhombic, pseudo-cubic crystal structure of ZrO$_2$ is typically stabilized under very high-temperature or strain conditions. This indicates that the chemical strain of the host matrix may favor the formation of the orthorhombic phase of ZrO$_2$. 

(ii) Second, we also observe a prominent peak at ${Q = 2.24~\AA^{-1}}$ ($2\theta = 36.6\degree$), which gradually diminishes with increasing Zr concentration from $\sim$3.6 at\% to 10.6 at\% (Fig.~3(b)). This peak position closely matches several possible phases: (a) two common secondary phases that generally form during Nb$_3$Sn synthesis, Nb$_6$Sn$_5$(026) and NbSn$_2$(422); and (b) NbC(004). This observation suggests that the secondary phases Nb$_6$Sn$_5$, NbSn$_2$, or NbC tend to diminish with increasing Zr concentration.
The case of $\sim$0.6 at\% Zr is anomalous, as it shows the maximum intensity of this peak. We infer that this may be due to higher carbon and contaminant incorporation in the material, since Zr has a strong affinity for both O and C atoms. During low-power deposition, only the very surface layers of the Zr target are sputtered, which may lead to a higher impurity concentration. We can correlate this result with our observation of the high hydrocarbon and NbC content in the sample with $\approx$0.6 at\% Zr using X-ray photoemission spectroscopy.

(iii) The third interesting observation is the strong out-of-plane orientation of Nb along the (110) direction. This indicates that the Nb buffer layer exhibits oriented growth along the out-of-plane direction. To explore this further, we performed pole-figure measurements, where the angle $\psi$ was varied from 0$\degree$ to 75$\degree$, while $\phi$ was varied from 0$\degree$ to 360$\degree$, with the Bragg condition satisfied for the asymmetric Nb(200) plane, as shown in Fig.~2(d).
\begin{figure*}
    \centering
    \includegraphics[width=1.1\linewidth]{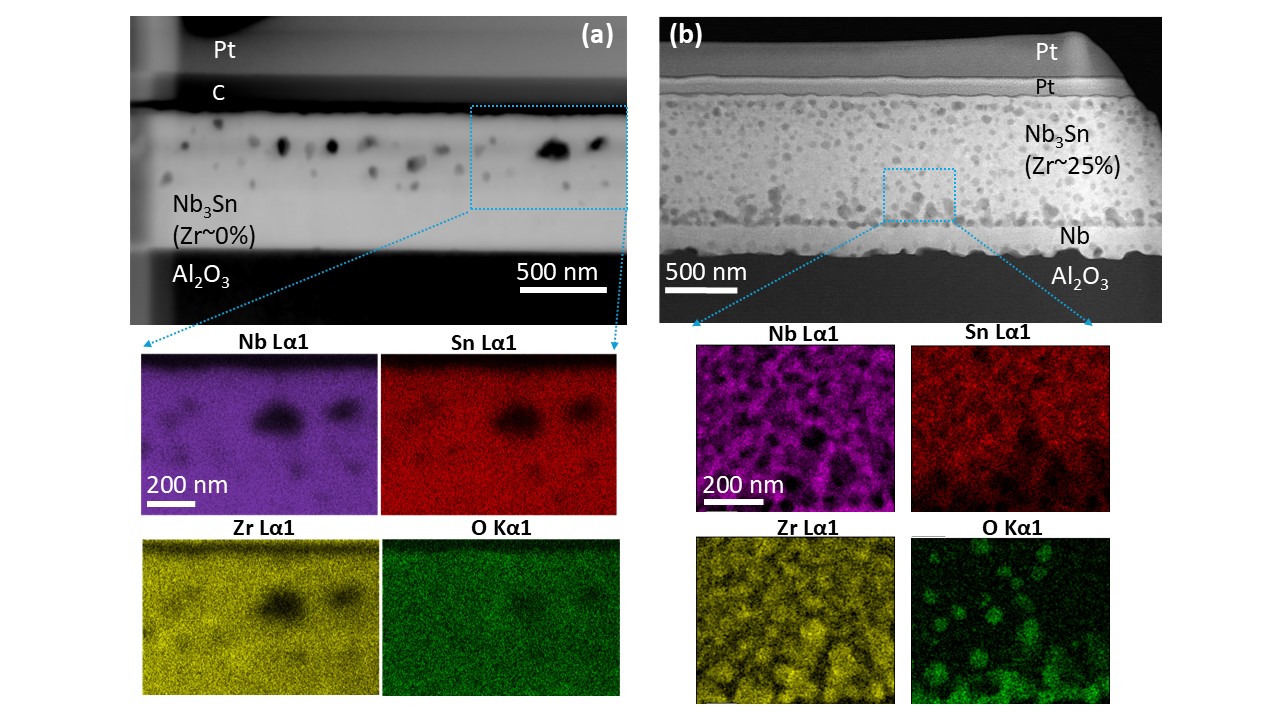}
    \caption{Cross-sectional HAADF-STEM images of Nb$_3$Sn films with (a) $\approx$0~at.\% Zr and (b) $\approx$25~at.\% Zr, with the corresponding EDS maps (Nb~L$\alpha$, Sn~L$\alpha$, Zr~L$\alpha$ and O~K$\alpha$) of the regions marked by the dashed rectangles. (a) All signals decrease across the large irregular regions, identifying them as voids; the contrast in the Zr~L$\alpha$ map arises from overlap with the Nb~L$\alpha$ line. (b) The Zr and O signals increase and the Nb and Sn signals decrease across the dark regions, identifying them as ZrO$_2$ precipitates.}
    \label{fig:enter-label}
\end{figure*}

We observe that the Nb buffer layer is textured with a six-fold symmetry, following the hexagonal structure of the Al$_2$O$_3$ substrate (Fig.~3(f)). The distinct spots representing the distribution of planes become more diffuse when the Zr concentration is increased beyond 3.6 at\%. This experiment provides a useful range of Zr concentration in Nb$_3$Sn grown on Al$_2$O$_3$ substrates over which the epitaxial relationship between the substrate and thin-film layer can be preserved.

\subsection{Evolution of surface morphology and bulk homogeneity of Nb$_3$Sn with increasing Zr content}
To understand how the bulk and surface morphology of Nb$_3$Sn change with Zr content, cross-sectional STEM images of samples with Zr concentrations ranging from 0 to 24.8~at.\% were compared. Figure~3(a) shows a cross-sectional HAADF-STEM image of a Nb$_3$Sn sample with $\approx$0~at.\% Zr. Large irregular regions are observed, mostly within the upper 500~nm of the film, with an average size of $128 \pm 28$~nm (range 105--184~nm) and an average area of $(1.1 \pm 0.6) \times 10^{4}$~nm$^2$. The contrast of these regions varies, which can be attributed to their different depths within the TEM lamella. Smaller regions, similar to the ZrO$_2$ precipitates discussed in the previous section, are also present, although they are less clearly resolved in this image. To identify the large irregular regions in the upper half of the film, STEM-EDS was performed on the rectangular region of interest marked in Fig.~3(a). The Nb, Sn and O signals decrease significantly across these regions, indicating that they are most likely voids. Large irregular voids of similar size are also observed in the sample with $\approx$2.6~at.\% Zr, but not in samples with higher Zr concentrations. Figure~3(b) shows a cross-sectional HAADF-STEM image of a sample with $\approx$24.8~at.\% Zr, which contains considerably smaller dark regions with an average size of $46 \pm 24$~nm (range 14--161~nm) and an average area of $(1.7 \pm 1.9) \times 10^{3}$~nm$^2$. The EDS elemental maps of the rectangular region of interest in Fig.~3(b) confirm that these dark regions are ZrO$_2$ precipitates.

SEM images of the surface of Nb$_3$Sn films with different Zr concentrations are shown in Fig.~4. The void fraction and void dimensions were determined from the SEM images by defining voids as regions with a grey level below 60\% of the median grey level of the image. In the film with $\approx$0~at.\% Zr [Fig.~4(a)], grains with similar growth orientations form clusters, and voids appear at the sharp boundaries between these clusters (inter-cluster voids). These voids occupy approximately 18\% of the imaged surface, with a number density of $\approx$46~$\mu$m$^{-2}$. The irregular inter-cluster voids have an average size of $81 \pm 47$~nm and an average area of $(3.9 \pm 4.0) \times 10^{3}$~nm$^2$, with the largest voids reaching $\approx$2.9 $\times 10^{4}$~nm$^2$.

In the film with $\approx$2.6~at.\% Zr [Fig.~4(b)], the grains are still clustered, but the voids cover only $\approx$2.7\% of the imaged surface, with a number density of $\approx$14~$\mu$m$^{-2}$, and their average size and area decrease to $51 \pm 25$~nm and $(1.9 \pm 2.0) \times 10^{3}$~nm$^2$, respectively. In addition, small particles appear on the surface. EDS confirms that these surface particles are ZrO$_2$ precipitates with dimensions similar to those observed in the cross-sectional images.

In the film with $\approx$2.6~at.\% Zr [Fig.~4(b)], the grains remain clustered, but the void fraction decreases to $\approx$2.7\% of the imaged surface and the number density to $\approx$14~$\mu$m$^{-2}$. The voids are also smaller, with an average size of $51 \pm 25$~nm and an average area of $(1.9 \pm 2.0) \times 10^{3}$~nm$^2$. In addition, small particles appear on the surface, which EDS identifies as ZrO$_2$ precipitates with dimensions similar to those observed in the cross-sectional images.

As the Zr concentration increases from 10.6 to 19.2~at.\% [Figs.~4(c) and 4(d)], the number density of voids decreases gradually, and above 20~at.\% Zr [Figs.~4(e) and 4(f)], it becomes negligible. At these Zr concentrations, the ZrO$_2$ precipitates also disappear from the surface, suggesting that they form below the surface when the Zr concentration exceeds 20~at.\%.

\begin{figure*}[hbt]
    \centering
    \includegraphics[width=1\linewidth]{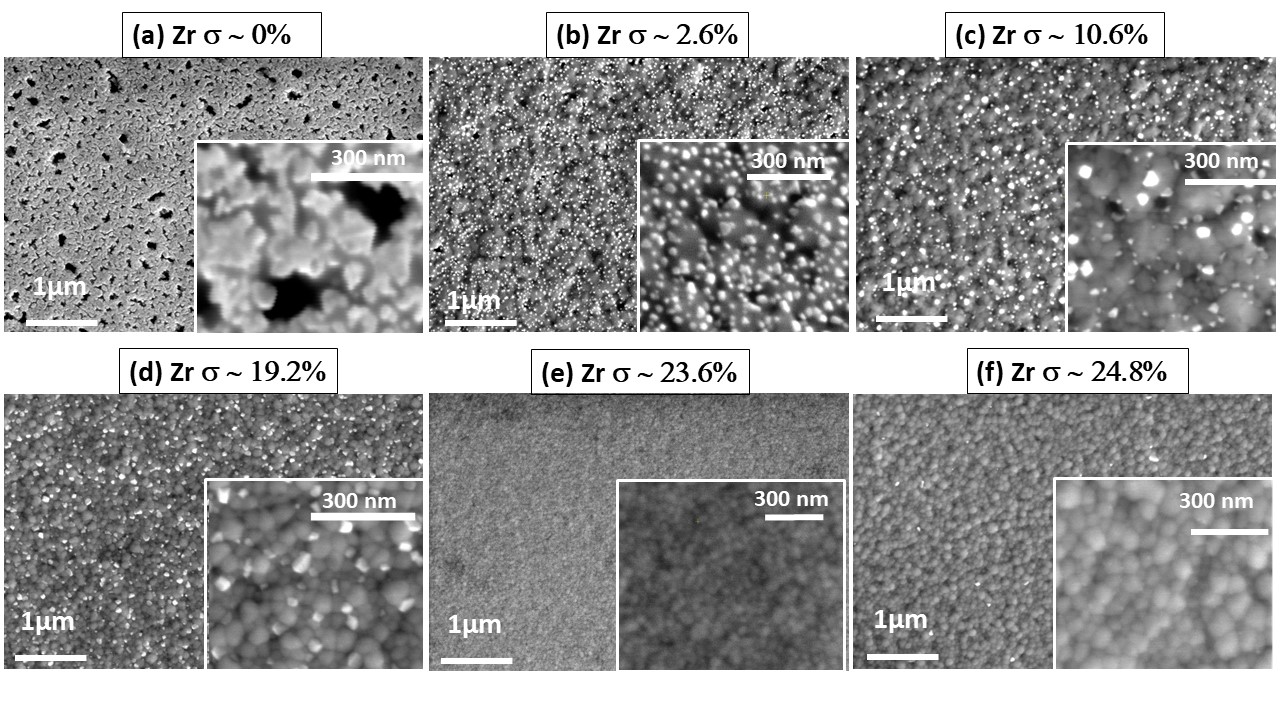}
    \caption{Secondary-electron SEM images (scan area $4~\mu\mathrm{m} \times 3~\mu\mathrm{m}$) showing the evolution of the surface morphology of Nb$_3$Sn films with Zr content: (a) 0, (b) 2.6, (c) 10.6, (d) 19.2, (e) 23.6 and (f) 24.8~at.\% Zr. The insets show $0.6~\mu\mathrm{m} \times 0.6~\mu\mathrm{m}$ areas.}
\end{figure*}

Surface depressions such as voids and grooves locally enhance the RF magnetic field, which lowers the field at which vortices penetrate and increases RF dissipation\cite{Kubo2013, Lechner_2025, Harbick2025}. The incorporation of Zr into sputtered Nb$_3$Sn coatings may therefore help reduce the overall RF power dissipation.

\begin{figure*}[hbt]
    \centering
    \includegraphics[scale=0.5]{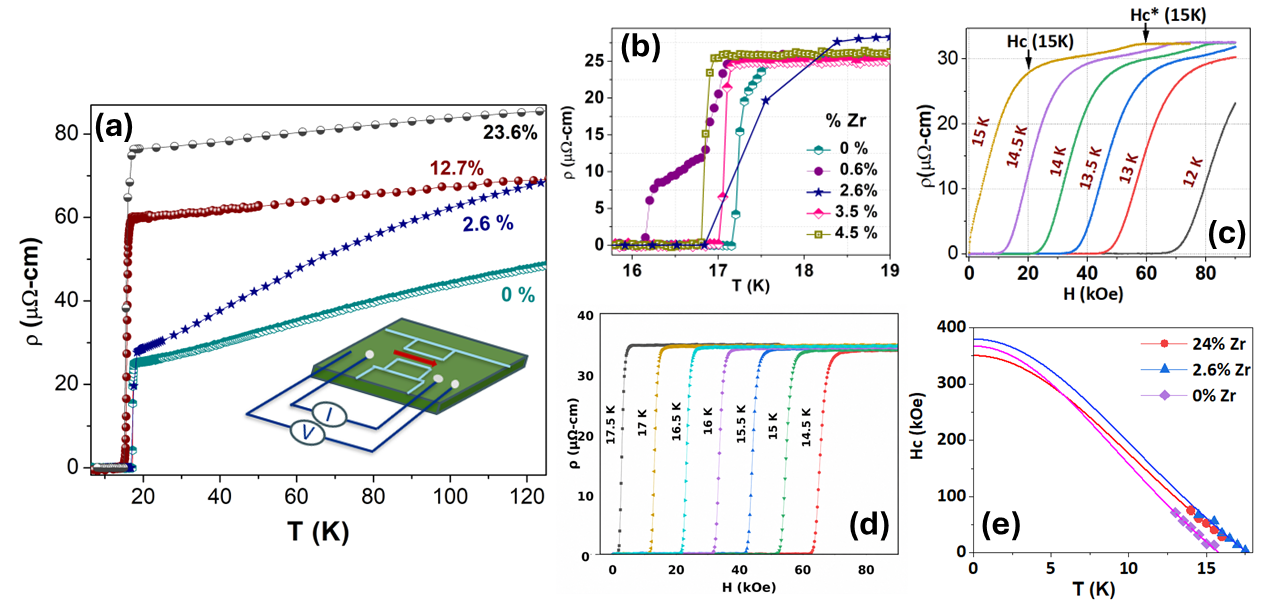}
    \caption{(a): Resistivity as a function of temperature for Nb$_3$Sn films with different Zr concentrations, the schematic illustrates the FIB defined current channels for transport measurements shows.(b) The variation of $\rho (T)$ curves in the close vicinity of superconducting transition temperature for Zr concentration varying from 0-4.5 at\%. (c, d) $R(H)$ curves for samples with 0.6 at\% Zr and 23.6 at\% Zr, respectively, measured at temperatures $T \leq T_C$. (e) Measured $H_{C2}$ values as a function of temperature, along with extrapolation of the data points to estimate the 0 K value of $H_{C2}$.}
    \label{fig:enter-label}
\end{figure*}

\subsection{Effect on superconducting properties of Nb$_3$Sn with Zr inclusions}

To characterize the superconductivity of Nb$_3$Sn with Zr inclusions, we measured the superconducting transition temperature and upper critical magnetic field (H$_{C2}$) by measuring the resistivity as a function of temperature, $R(T)$, and magnetic field, $R(H)$. For both $R(T)$ and $R(H)$ measurements, we first fabricated a well-defined channel of $1~\mathrm{mm} \times 200~\mu\mathrm{m}$ for current transport using focused-ion beam milling, as illustrated in the schematic shown inside Fig.~5(a). The current-transport channel was defined by ion-milling through the $\sim 1~\mu\mathrm{m}$ Nb$_3$Sn-Zr layer and $\sim 200~\mathrm{nm}$ of Nb to reach the insulating Al$_2$O$_3$ substrate layer. This channel fabrication is useful for current-transport measurements because it allows the use of a very small excitation current, which prevents sample overheating and helps measure the transition temperature accurately. The narrow width of $200~\mu\mathrm{m}$ helped to record a low noise signal even with a small current of $20~\mu\mathrm{A}$, while the $1~\mathrm{mm}$ long channel length provides an extended current path.

The $R(T)$ curves for Nb$_3$Sn samples with different Zr concentrations, ranging from 0 at\% to 23.6 at\%, are shown in Fig.~5(a). The extracted characteristic superconducting parameters are listed in Table~I. The room-temperature normal-state resistivity of undoped Nb$_3$Sn is observed to be $\rho(300~\mathrm{K}) = 69.1~\mu\Omega\cdot\mathrm{cm}$. We observe a noticeable increase in $\rho(300~\mathrm{K})$ and a decrease in RRR for Zr concentrations of samples representing high atomic concentrations of Zr: 12.7 at\% and 23.6 at\%. However, for lower Zr concentrations, non-monotonic behavior is found for both $\rho(300~\mathrm{K})$ and RRR. No sign of a chemical strain induced maternistic transition was observed over the studied range of Zr inclusion in Nb$_3$Sn. TThis behavior can be understood from the fact that even the lowest measured $\rho(300~\mathrm{K})$ value remains higher than the theoretically predicted critical normal-state resistivity \cite{Strain_Nb3Sn, godeke2006review}.

\begin{table}
    \centering
    \begin{tabular}{c|c|c|c|c|c|c}
      Zr &   T$_c$($\rho$) &T$_c$(m)  & $\Delta T_c (m)$  & $\rho _{300K}$  & RRR  & H$_{C2} (0 K)$ \\
      \% &  (K) &  (K) & (K)& $\mu\Omega-cm$  & $\rho_{300K}/\rho_{20K}$  &  kOe \\
      \hline
      \hline
         0 & 17.5 & 17.0 & 0.9  & 69.1  & 2.7& 268.4 (14)   \\
         0.6 & 17.3, 16.3 & 16.5 & 1.2 & 60.4  & 2.5 & 277.1(37) \\
         2.6 & 17.8  & 17.7  & 1.2  & 73.2 & 3.4 & 302.5 (52) \\
         3.6 & 17.4 &17.1  & 1.3  & 64.8 & 2.6 & 297.9(19)\\
         4.5& 17.1  & 16.7 & 1.9 &  61.9 & 2.3 & 305.1( 41) \\
         12.7& 16.8 & 16.4 & 2.4  & 77.7 & 1.3 & 308.7 (39)\\
         23.6 & 16.6 & 16.3 & 3.9  & 94.8 & 1.2 & 288.4 (26) \\
    \end{tabular}
    \caption{ Characteristic superconducting parameters for variable Zr in Nb$_3$Sn}
    \label{tab:placeholder}
\end{table}
\begin{figure*}[hbt]
    \centering
    \includegraphics[scale=0.5]{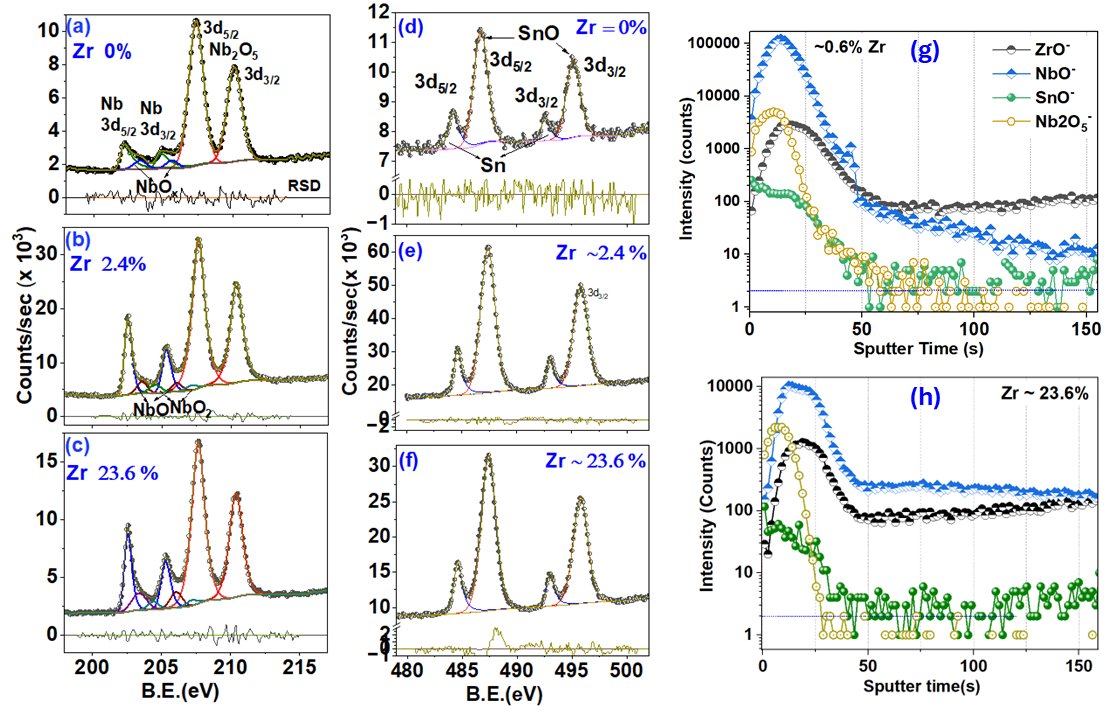}
    \caption{ (a-c): Nb core level XPS 3d spectra of with various Zr concentrations (a) 0 at\%, (b) 2.4 at\% (c) 23.6 at\%; (d-f) Sn core level 3d specta with Zr concentrations (d) 0 at\%,(e) 2.4 at\% and (f) 23.6 at\%  (g,h) SIMS depth profiles of samples with Zr concentration of 0 at\% and 23.6 at\%, respectively.  }
    \label{fig:enter-label}
\end{figure*}
The temperature-dependent resistivity in the close vicinity of resistive superconducting transition, $T_C(\rho)$, is shown in Fig.~5(b). The value of $T_C(\rho)$ is determined from the intersection of two extrapolated straight-line fits to the regions before and after the higher-temperature side of the superconducting transition. The temperature-dependent bulk magnetic moment was also measured in zero-field-cooled (ZFC) mode under an applied magnetic field of $\mu_0H = 0.01~\mathrm{T}$ to determine the magnetic superconducting transition temperature, $T_C(m)$. These data are presented in the Supplementary Material (Fig.~S4). While $T_C(\rho)$ represents the superconducting transition along the least-resistive current path, $T_C(m)$ provides a measure of the bulk superconducting response.

The resistive superconducting transition temperature, $T_C(\rho)$, of undoped Nb$_3$Sn is measured to be $17.5~\mathrm{K}$, which closely matches values reported for state-of-the-art magnetron-sputter-grown Nb$_3$Sn materials~\cite{SAYEED2021148528,SAYEED2023112019}. The value of $T_C(\rho)$ does not change substantially for Zr concentrations in the range of 0-4.5 at\%; however, at higher Zr concentrations, it starts to  decrease gradually with increment in Zr at.\%. For one sample with Zr 2.6 at\%, we observed a slight increase in both $T_C(\rho)$ and $T_C(m)$ values. For all the measured samples, both $T_C(\rho)$ and $T_C(m)$ show similar and consistent variations with increasing Zr fraction. The width of the magnetic superconducting transition, $\Delta T_C(m)$, is defined as the temperature span over which the magnetic signal changes from 90 at\% to 10 at\% of the full transition step. We observe that $\Delta T_C(m)$ remains close to the value of undoped Nb$_3$Sn below a Zr fraction of 4.5 at\%, and starts to broaden as the Zr inclusion increases.
An interesting observation is noted for the sample with 0.6 at\% Zr, which shows two distinct, closely spaced superconducting transitions: one at $16.3~\mathrm{K}$ and another at $17.2~\mathrm{K}$. The presence of two transitions indicates the possibility of two distinct Nb$_3$Sn phases. However, the measured $T_C$ values do not correspond to any known secondary phase formed during Nb$_3$Sn synthesis or to niobium surface oxides~\cite{Godeke_2006}. The closely spaced transitions can therefore be attributed to surface Nb$_3$Sn states with slightly different atomic coordination environments compared with the bulk Nb$_3$Sn phase~\cite{yang2016,Mentink_2017}..
Combining this observation with our XRD results, we infer that a small, optimal amount of Zr inclusion favors stabilization of the stoichiometric Nb$_3$Sn A15 structural phase. However, when the Zr concentration increases above this optimal level, ZrO$_2$ clusters start to form phase-separated ordering, which further reduces the superconducting transition temperature, broadens the thermal transition width, and increases the $\rho(300~\mathrm{K})$ values.

Figures~5(c) and 5(d) show the resistivity as a function of applied magnetic field, measured at selected temperatures for Nb$_3$Sn with 0.6 at\% and 2.6 at\% Zr, respectively. The two superconducting phases in Nb$_3$Sn with 0.6 at\% also show two different values of upper critical magnetic field (H$_{C2}$). However, only one distinct value of (H$_{C2}$) is observed for other Zr concentrations. To estimate the value of H$_{C2}$ at 0 K, the temperature variation of H$_{C2}$ is simulated and extrapolated. To estimate the value of  H$_{C2}$ at 0 K, the temperature variation of H$_{C2}$ is simulated and extrapolated based on the Werthamer-Helfand-Hohenberg (WHH) equation\cite{WHH}:

$$ H_{C2}(T)= H_{C2}(0)[1-(T/T_C)^\lambda]$$
where the value of $\lambda$ is experimentally measured to be $\approx$ 1.52. This model incorporates the electron-spin and spin-orbital scattering and provides reasonable agreement with a wide range of commercial and experimental Nb$_3$Sn wires\cite{GLAG,Taylor_2005}. The absolute values of upper critical magnetic field for Nb$_3$Sn sample with 0 at\%Zr is 268(13) kOe, which closely matches with the H$_{C2}$ value of sputter-coated Nb$_3$Sn.With increasing Zr concentration, we statistically observe an overall enhancement in the $H_{C2}$ value; however, some data points show slight non-monotonicity. 

\subsection{ Effect on nature of surface oxides of Nb$_3$Sn}

To evaluate how the surface oxides of Nb and Sn are modified by Zr inclusion, we recorded high-resolution core-level XPS spectra of Nb-3d, Sn-3d, Zr-3d, O-1S and C-1S. Fig.6 shows the core 3d level XPS profiles for Nb and Sn for different concentrations of Zr inclusion. The spectra in Fig.6(a) represents the Nb-3d core level XPS profile for Nb$_3$Sn with 0 at\% Zr. The peak profile is fitted with three components and their spin-orbit coupled counterparts: Primary Nb oxide, Nb$_2$O$_5$ at 207.7 eV ($\Delta_{SOC}$ = 2.8eV) with Gaussian-Lorentzian(GL) line shape, non-stoichiometric sub-oxide NbO$_x$ at 203.6 eV ($\Delta_{SOC}$= 2.6 eV) with  GL line shape, and Nb metal signal at 202.1 eV($\Delta_{SOC}$ = 2.7 eV) with asymmetric Lagrangian (LA) function as line-shape. The XPS spectra corresponding to 2.4 at\% Zr and 23.6 at\% [Fig. 6(b) and (c)] are also fitted with similar Nb$_2$O$_5$ and Nb-metal peak profiles, however, two distinct sub-oxides signals are now incorporated as: NbO at 203.3 eV ($\Delta_{SOC}$= 2.6 eV) and NbO$_2$ at 204.5 eV ($\Delta_{SOC}$= 2.7 eV) with GL line profiles.

If we compare the Nb-3d peak-profiles of samples with different Zr concentration, we clearly see a relative enhancement in Nb-metal signal when we increased the Zr concentration from 0 at\% to 2.4 at\%. Quantitatively, the Nb metal signal in 3d-spectra increases from 12.8 at\% to 20.7 at\%  and Nb$_2$O$_5$ signal reduces from 82.8 at\% to 71.6 at\% when we increase the Zr from 0 at\% to 2.4 at\%. However, there is no significant change when we increase the Zr concentration further from 2.4 at\% to 23.6 at\%. This observation indicates that increasing Zr concentration possibly prevents the oxygen diffusion and leads to thinner formation of primary surface oxide Nb$_2$O$_5$. Table.II describes the peak-intensity to atomic-subshell photoionization cross-section ratios for elements and oxides.  Perkins et. al,\cite{Perkins1977} studied the thermal evolution of oxygen diffusion in Nb-Zr alloys and noted that oxygen diffusivity decreases from 8.08 to 3.03 cm$^2$/s x 10 $^9$ when we increase the Zr concentration from 0.25 at\% to 1 at\%. Our results confirm the similar behavior of oxygen diffusivity in Nb$_3$Sn. 
\begin{table}[!htbp]
\centering
\setlength{\tabcolsep}{6pt}
\renewcommand{\arraystretch}{1.1}
\begin{tabular}{c|ccc}
Zr (\%) & $\Theta$(Nb/Nb$_2$O$_5$) & $\Theta$(Nb/Sub-oxides) & $\Theta$(Nb/Sn) \\
\hline
\hline
0    & 0.14 & 2.6  & 0.23 \\
2.6  & 0.28 & 4.3  & 0.43 \\
23.6 & 0.29 & 1.84 & 0.64 \\
\end{tabular}
\caption{Peak-intensity to atomic-subshell photoionization cross-section ratios for elements and oxides.}
\label{tab:oxide_ratio}
\end{table}

Fig. 6(d-f) display how Sn surface oxides vary with increasing Zr concentration. Sn-3d core level profiles consist of two main features: Primary surface oxide SnO$_2$ or SnO at 486.9eV ($\Delta_{SOC}$= 2.7eV) and Sn metal at 484.7eV ($\Delta_{SOC}$= 8.4eV). The Binding energy difference between SnO$_2$ and SnO is very small (0.2-0.3eV) and we are not able to distinguish or resolve in between these two oxides. When we compare the peak profiles for samples with different Zr concentrations, we observe that Sn oxides and metal ratio are not much affected by increasing the Zr concentration. Additionally, we also recorded the Zr 3d core level spectra (shown in Supplementary materials, Fig. S2), which indicates the presence of small Zr metal cluster on the surface for the sample which has 23.6 at\% Zr. 
Fig. 6(g \& h) presents the background-subtracted ToF-SIMS depth profiles of the primary and secondary surface oxides of Nb$_3$Sn samples with 0 at\% and 23.6 at\% Zr, respectively. Comparing the Nb$_2$O$_5$ profiles for these samples, it is clear that the sputter time required for the Nb$_2$O$_5$ signal to reach the noise level in the Nb$_3$Sn sample with 0 at\% Zr is approximately twice that of the Nb$_3$Sn sample with 23.6 at\% Zr. This observation is also consistent with the results obtained from XPS. Based on previous studies\cite{Sam_PhysRevApplied.13.014024, ito:srf2021-frofdv01}, it is expected that the decomposed layer of primary surface oxide helps in reducing the surface resistance and enhancement in intrinsic Quality factor.

\section{Conclusions}
The material properties of Nb$_3$Sn were modified by introducing a tunable fraction of Zr in Nb$_3$Sn host matrix via co-sputtering process. The combined co-sputtering and post-deposition annealing treatment lead to Nb$_3$Sn phase incorporated with ZrO$_2$ precipitates of average dimensions ranging from 20-100 nm. There are three important conclusions from this study: firstly, we noted that the density of the surface and bulk voids, as well as their average dimensions can be reduced by increasing the Zr content in Nb$_3$Sn; secondly, we also observed that increasing Zr concentration up to the optimal level can moderately improve the value of superconducting transition temperature and significantly enhance the upper critical magnetic field; thirdly, increasing the Zr concentration reduces oxygen diffusion in Nb, resulting in a thinner formation of primary Nb surface oxides. These results indicate that inclusion of Zr in Nb$_3S$n sputtered coating is a promising method to improve the material quality of Nb$_3$Sn based SRF cavities. 
%and will help to reduce the overall RF power dissipation of Nb$_3$Sn based SRF cavities. 

\section{Acknowledgment}
Authors are thankful to Zuhawn Sung, Adam Clairmont, and Brad Tennis for their help and technical support. This work was produced by Fermi Forward Discovery Group, LLC under Contract No. 89243024CSC000002 with the U.S. Department of Energy, Office of Science, Office of High Energy Physics. Publisher acknowledges the U.S. Government license to provide public access under the DOE Public Access Plan DOE Public Access Plan. Part of this research has been sponsored by the Laboratory Directed Research and Development Program of Fermi National Accelerator Laboratory, managed by Fermi Research Alliance, LLC for the U.S. Department of Energy. Work at Old Dominion University was supported by U.S. Department of Energy, Office of Science, Office of Accelerator R\&D and Production, under contract No. DE-SC0022284.                   

\bibliography{apssamp}

%\newpage
%\Section{Supplementary Material}

%\subsection{Preferred distribution patterns of ZrO$_2$ in Nb$_3$Sn}

%\begin{figure*}
%    \centering
%   \includegraphics[width=1\linewidth]{S1.jpg}
%    \caption{Annular Dark Field cross-sectional STEM images of Nb$_3$Sn with different Zr concentrations; (a) with approximately 10.8\% Zr (b) with approximately 23.6\% Zr }
%    \label{fig:enter-label}
%\end{figure*}
\end{document}